\documentclass[conference]{IEEEtran}
\usepackage{cite}
\ifCLASSINFOpdf
   \usepackage[pdftex]{graphicx}
\else
\fi
\usepackage[cmex10]{amsmath}
\usepackage{algorithmic}
\usepackage{url}
\begin{document}
\title{Are all Social Networks Structurally Similar? A Comparative Study using Network Statistics and Metrics}

\author{\IEEEauthorblockN{Aneeq Hashmi}
\IEEEauthorblockA{National University of Science and Technology\\
Karachi, Pakistan\\
Email: aneeq.hashmi@nu.edu.pk}
\and
\IEEEauthorblockN{Faraz Zaidi}
\IEEEauthorblockA{Karachi Institute of Economics and Technology\\
Karachi, Pakistan\\
Email: faraz@pafkiet.edu.pk}
\and
\IEEEauthorblockN{Arnaud Sallaberry}
\IEEEauthorblockA{University of California, \\
Davis, USA\\
Email: asallaberry@ucdavis.edu}
\and
\IEEEauthorblockN{Tariq Mehmood}
\IEEEauthorblockA{National University of Science and Technology\\
Karachi, Pakistan\\
Email: tariq.mehmood@nu.edu.pk}}


%Telephone: (800) 555--1212, Fax: (888) 555--1212}


\maketitle

\begin{abstract}

The modern age has seen an exponential growth of social network data available on the web. Analysis of these networks reveal important structural information about these networks in particular and about our societies in general. More often than not, analysis of these networks is concerned in identifying similarities among social networks and how they are different from other networks such as protein interaction networks, computer networks and food web.

In this paper, our objective is to perform a critical analysis of different social networks using structural metrics in an effort to highlight their similarities and differences. We use five different social network datasets which are contextually and semantically different from each other. We then analyze these networks using a number of different network statistics and metrics. Our results show that although these social networks have been constructed from different contexts, they are structurally similar. We also review the snowball sampling method and show its vulnerability against different network metrics.

\end{abstract}


% \fi
\IEEEpeerreviewmaketitle

\section{Introduction}

The web has provided us with a platform to build huge social networking webistes\cite{lewis08} and communication channels~\cite{boyd07} with hundreds and thousands of users. These networks provide challenging opportunities for researchers to analyze and explore how virtual societies exist in the cyber world and how they impact our societies in the real world~\cite{acar08}. Many useful applications for these online networks have been found such as information diffusion~\cite{iribarren11}, corporate communication~\cite{scott05} and supplier-customer relationship~\cite{hu05}. 

Often these social networks are compared to other real world networks such as protein interaction networks\cite{cannataro10}, computer networks~\cite{wellman01} and food web~\cite{williams00}. For example, Newman studied the property of assortativity~\cite{newman03b} only present in social networks where individuals of similar degree have the tendency to connect to each other. Not much attention has been given to the differences and similarities of contextually and semantically different social networks. As social networks can take many different forms depending upon how they are constructed based on the context and semantics associated with the data~\cite{rosen11} and hence poses the question of whether different social networks have the same network structure. 

%When compared with other social networks, often the studies consider only 2 to 4 datasets and/or are compared using only limited network metrics.

%This study presents an opportunity to critically analyze how similar social networks are, and to what extent does their behavior with respect to different network metrics overlap.

In this paper, we address this question and try to answer it empirically. We use five different social network datasets and compare them using different network statistics and metrics. Our results show high similarity among structural behavior of these networks with only slight differences. Major contributions include highlighting structural similarities and dissimilarities among different social networks. We also review different network sampling methods and focus on the most widely accepted snowball sampling method. Our experiments show that this method does not always produce correct samples in terms of structural properties of a network and one should be careful when drawing conclusions when this method is used.

The rest of the paper is organized as follows: In the next section, we review the literature where online social networks have been analyzed. Section \ref{sec::data} describes the data sets used for experiemtation. In sections \ref{sec::stats} and \ref{sec::network}, we review a number of network statistics used for comparative study of networks. Section \ref{sec::experimental} describes how the samples were collected and the shortcomings of the snowball sampling method. We comparatively analyze different networks in section \ref{sec::inferences} and finally we draw conclusion and discuss future research prospects in section \ref{sec::inferences}.

\section{Related Work}\label{sec::related}

\subsection{Analysis of Online Social Networks}

Jacob Moreno's~\cite{moreno34} seminal work on runaways from the Hudson school for girls gave birth to sociometry. Since then, this field has grown steadily. Recent interest in this field was triggered by the work on small world ~\cite{watts98} and scale free networks~\cite{barabasi99}. Further thrust to this field was given by the availability of large size social network data from online sites such as Facebook and Twitter. Since then, many researchers have actively pursued research in social network analysis (SNA) mainly due to its wide application in different fields of research ranging from genetics to nanoelectronics, from disease epidemics to product marketing. We briefly review some literature related directly to using online social network data.

Garton et al.~\cite{garton97} emphasized that earlier, research effort concentrated on studying how people use computers to communicate (computer mediated communication) rather than studying the social networks generated by this medium. They describe methods to identify sources to collect and analyze social network data focusing on how online communication systems provide a perfect platform to study virtual communities and interaction networks. 

Kumar et al.~\cite{kumar06} study the structural evolution of large online social networks using \textit{Flickr} and \textit{Yahoo! 360} data sets. The authors show the presence of `stars' in both these networks concluding that both the graphs are qualitatively similar.

Ahn et al~\cite{ahn07} compare the structure of three online social networks: \textit{Cyworld}, \textit{MySpace}, and \textit{Orkut}. They observe a multi-scaling behavior in Cyworld's degree distribution and that the scaling exponents of MySpace and Orkut are similar to those from different regions in the Cyworld data. They also validate the snowball sampling on Cyworld using degree distribution, clustering coefficient, degree correlation (also known as assortativity)~\cite{newman03a} and average path length. 

Mislove et al.~\cite{mislove07} use \textit{Flickr}, \textit{LiveJournal}, \textit{Orkut}, \textit{YouTube} using degree distribution, in-degree and out-degree, average path length, radius, diameter and assortativity metrics. Their analysis shows that social networks differ from other networks as they exhibit much higher clustering coefficient. They also show that social network have a higher fraction of symmetric links.

Lewis et al.~\cite{lewis08} investigate \textit{Facebook} data emphasizing five distinct features. First, the correctness of data is ensured as it is downloaded from the internet avoiding classical problems of interviewer effect~\cite{campbell86}, imperfections in recall~\cite{brewer00} and other measurement errors. Second, the dataset is complete as it contains information about all the existing social ties in the network. Third, the data is collected over four years allowing temporal analysis of the social dynamics taking place in the network. Fourth, data on social ties is collected for multiple social relations:  \textit{Facebook Friends}, \textit{Picture Friends} and \textit{Housing Friends}. Finally, with users providing data for their favourite music, movies and books: the dataset is quite rich and provides new research opportunities.

Benevenuto et al.~\cite{benevenuto09} use an entirely different approach to study and analyze social networks by studying the click streams generated when a user accesses a social network site. Four online social networks: \textit{Orkut}, \textit{MySpace}, \textit{Hi5} and \textit{LinkedIn} were used to collect data of 37024 users over a 12-day period. The authors studies patters such as how frequently and for how long people connect to these networks, and how frequently they visit other people's pages. They also compared the click stream data and the topology of the friends social network of Orkut. Results reveal publicly \textit{visible} social interactions such as commenting profiles as well as \textit{silent} social interaction such as viewing profile and photos.

Rejaie et al~\cite{rejaie10} study \textit{MySpace} and \textit{Twitter} with the intent of finding the active population of these networks. They develop a measurement technique using the numerical user IDs assigned to each new user and the last login time of each user. This in turn helps to identify short lived users on the site and are termed as \textit{tourists}. Results show that the number of active users in these networks is an order of magnitude smaller than the total population of the network.

Interesting observations about online social networks can be found in \cite{howard08}. More comprehensive and recent review of literature on social networks can be found in \cite{borgatti09,scott11}

\subsection{Network Statistics and Metrics}

There are a number of network statistics and metrics in the literature. A detailed description of the metrics we have used is given in section~\ref{sec::stats} and section~\ref{sec::network}. We consider node metrics that are widely used in the research community, or the most representatives ones as these basic metrics have been used to derive new variants. For example, we have used \textit{betweenness centrality}\cite{freeman77} instead of \textit{stress centrality}\cite{shimbel53} which simply counts the absolute number of shortest paths.  

Another criterion of selecting the metrics we have used is that they are all applicable on undirected networks. For example, Burt's constraint~\cite{burt05} to calculate the local cohesiveness is calculated for directed graphs only. although some of the networks that we are using are directed in nature, but we limit our study to only undirected graphs.

An important class of networks that we have not considered in this study is the metrics calculated on edges. A good resource to review these metrics is the book by Brandes and Erlenach~\cite{brandes05}.
%** add more here

%Howard ~\cite{howard08}

%,martino06,jamali06}

%Jamali and Abolhassani~\cite{jamali06} consider web as a social network 

%"We review some basic concepts of social network analysis, describe how to collect and analyze social network data, and demonstrate where social network data can be, and have been, used to study computer-mediated communication. Throughout, we show the utility of the social network approach for studying computer-mediated communication, be it in computer-supported cooperative work, in virtual community, or in more diffuse interactions over less bounded systems such as the Internet."

\section{Data Sets}\label{sec::data}

We have used a number of different data sets representing a variety of social networks used for analysis by the research community. The data sets are described below:

\textbf{Twitter Friendship Network:} Twitter is one of the most popular social networks in the world. A friendship network is extracted by crawling the twitter database using the api\footnote{\url{api.twitter.com}}. Given a single user, the api returns a list of all the friends of the given user. We recursively applied this method to gather data of 2500 users starting from a single user. The complete network has 22002 edges.

\textbf{Epinions Social Network:} This is a \textit{who-trust-whom} online social network of a customer analysis site Epinions.com\footnote{\url{http://www.epinions.com/}}. Members of the site can either agree or disagree to trust each other. All the reliable contacts interact and form a of Trust which is then shared with users on the basis of review ratings. We have downloaded this data from the stanford website\footnote{\url{http://snap.stanford.edu/data/}} where it is publicly available in the form of a text file. The network contains 75879 nodes and 508837 edges.

%\textbf{Live Journal Social Network:} Live Journal\footnote{\url{http://www.livejournal.com/}} is an online network with around 10 million members, majority of whom are extremely active. LiveJournal allow members to maintain journals, blogs and people are permitted to declare which other members are their associates. The network contains 4847571 nodes and 68993773 edges and is available on the stanford  website.

\textbf{Wikipedia Vote Network:} Wikipedia is a free encyclopaedia which is written collectively by assistants around the world. A small number of people are designated as administrators. Using the complete dump of Wikipedia page edit history, we selected all administrator elections and vote history data. Users are represented by nodes in the network and a directed edge from node \textit{i} to node \textit{j} represents that user \textit{i} voted on user \textit{j}. Again, the data is available from stanford website with 7115 nodes and 103689 edges.

\textbf{EU Email Communication Network:} This network was generated by using email data from a huge European research institution. Information was collected about all emails (incoming and outgoing) for a period of Oct 2003 to May 2005. Nodes represent email addresess and an edge between nodes \textit{i} and \textit{j} represents that \textit{i} sent at least one email to \textit{j}. The network contains 265214 nodes and 420045 edges and available from stanford.

\textbf{Author Network:} is a collaboration network of authors from the field of computational geometry. Two actors are connected to each other if they have co-authored an artifact together. The network was produced from the BibTeX bibliography obtained from the Computational Geometry Database `geombib', version February 2002\footnote{see~\url{http://www.math.utah.edu/~beebe/bibliographies.html}}. Problems with different names referring to the same person are manually fixed and the data base is made available by Vladimir Batagelj and Andrej Mrvar: Pajek datasets from the website\footnote{\url{http://vlado.fmf.uni-lj.si/pub/networks/data/}}. We only consider the biggest connected component containing 3621 nodes and 9461 edges.

All these five datasets model contextually and semantically different social relations from each other. Twitter network is a friend network and represents mutual acceptance from both individuals. Epinions network is similar in the sense that it requires mutual acceptance but differs as it requires a certain degree of trust rather than friendship. Wikipedia network is a directed network which represents the voting behavior of users to select administrators and is completely different from the previous two contexts. The fourth dataset is the Email network which is also a directed network where users are related to each other if a user has communicated to the other through email. Finally the Author network is an affiliation network~\cite{newman03} which are based on bipartite graphs and are related to each other by having an affiliation to a common research artefact.

%\textbf{Imdb Network:} is an actor-actor network where nodes represent actors and two actors are connected to each other if they have acted in a movie together. The data set we use here is a subset taken from the IMDB database (~\url{http://www.imdb.com/}) of movies. This network contains 7640 nodes and 277029 edges.

\section{Network Statistics}\label{sec::stats}

Table \ref{tbl::stats} shows some basic network statistics calculated on the above described data sets. We briefly define  these statistics below:

\textbf{Density} refers to the Edge-Node ratio of a network representing the average degree of a node in the network. \textbf{Highest Degree (HD)} is the highest node degree a node has in the network. \textbf{Diameter} is the number of edges on the longest path between any two nodes in the network. \textbf{Girth} of a graph is the path length of the shortest cycle possible. \textbf{Clustering Coefficient Global (CCG)} is the measure of connected triples in the network.  \textbf{Average Path Length (APL)} is the average number of edges traversed along the shortest paths for all possible pairs of network nodes. \textbf{Alpha($\alpha$)} is the constant obtained when a power-law distribution is fitted on the degree distribution of the network.

%This distance can be mathematically defined as:  $$APL=\frac{1}{n*(n-1)}\sum_{i,j} d_{ij} $$ where $d_{ij}$ is the geodesic distance from node \textit{i} to node \textit{j} and \textit{n} is the total number of nodes in the network.
%We used the igraph package available in the R software to calculate these values.

\textit{Density} values for Epinions and Wikipedia networks are comparatively very high representing high number of connections for each node in the network. High density of networks can be one reason for having high clustering coefficient for a network but in the presented datasets, the networks with the lowest density have the highest CCG values which represents an important structural trait for these network as they have slightly higher number of transitive triples. For the author network, this is inherent due to the construction method of the network as research artefacts with three or more than three authors will all form triads. This observation is more interesting for the email network where people exchange emails forming triads whereas relatively low values for the twitter network suggest that friend of a friend phenomena is not quite common when compared to the email network. Girth values of 3 for all these networks represents the presences of smallest possible cycle in the network.

The APL and $\alpha$ values of all the networks are quite close to each other again representing the similarity among the different networks. Low APL, High CCG and $\alpha$ values between 1.5 and 3 for twitter, email and author network represent the small world and scale free properties for these networks. The $\alpha$ value close to 1.2 for epinions and wikipedia network show a linear decay in the degree distribution and should not be classified as scale free networks. The histogram of degree distribution for all these networks is presented in Figure~\ref{fig::results}.

\begin{table}
\centering
\begin{tabular}{|l|c|c|c|c|c|}
\hline
& Twitter & Epinions & Wikipedia & Email & Author \\
\hline 
\hline
Nodes & 500 & 500 & 500 & 500 & 500\\
\hline 
Edges & 3099 & 13739 & 11672 & 2396 & 2404\\
\hline 
Density & 6.18 & 27.47 & 23.34 & 4.79 & 4.80\\
\hline
HD & 237 & 278 & 281 & 499 & 102\\
\hline 
Diameter & 11 & 7 & 12 & 7 & 10 \\
\hline
Girth & 3 & 3 & 3 & 3 & 3 \\
\hline
CCG & 0.19 & 0.43 & 0.35 & 0.54 & 0.60 \\
\hline 
APL & 2.6 & 1.93 & 2.10 & 1.98 & 2.87\\
\hline
$\alpha$ & 1.57 & 1.202 & 1.209 & 1.87 & 1.66\\
\hline
%Largest Clique & 5 &  & & 16 & 22 \\
%\hline
%Burt's Constraint & 0.27 & 0.066 & 0.067 & 0.512  & 0.399 \\
%\hline
\end{tabular}
\caption{Basic Statistics for the Data Sets used in Experimentation. HD= Highest Node Degree, CCG= Clustering Coefficient Global, APL=Avg. Path Length, $\alpha$=Power Law Fitting Constant} \label{tbl::stats}
\end{table}

\section{Network Metrics}\label{sec::network}

We use the following notation throughout this paper. A network is a graph $G(V,E)$ where $V$ is a set of nodes and $E$ is a set of edges. $u,v,w \in V$ represents nodes and $e \in E$ represents an edge in the network.

In this section, we briefly describe a number of network metrics frequently used in network analysis. All the metrics considered are node level metrics or can be derived for nodes. Metrics are grouped together into Element Level Centrality, Group Level Cohesion and Network Level Centrality metrics. The metrics we have considered for experimentation are most widely used metrics in network analysis but the list is certainly not complete and an exhaustive study remains part of our future work. 

\subsection{Element Level Centrality Metrics}
Element level metrics are calculated on individual elements of a graph, in this case for nodes. The term centrality refers to the idea where these elements are central in some sense in the graph.

\textbf{Degree} of node is an element level metric which refers to the number of connections a node has to other nodes. Degree distribution of nodes has been one of the most important metric of study for networks as the degree distribution of most real world networks follow power law~\cite{li05}.

%The metric works with undirected networks. The \textbf{Out-Degree} and \textbf{In-Degree} metrics are used for directed graphs where the former represents the number of connections a node has with others and the later represents the number of connections other nodes have with a particular node.

\subsection{Group Level Cohesion Metrics}
Group Level Metrics are metrics that are calculated for a small subset of nodes within the graph. The two metrics we consider here in our study are cohesion metrics that give a measure of how closely a group of nodes is connected to each other.

\textbf{Local Clustering Coefficient}~\cite{watts98} is a group level metric which counts the degree of connectedness among neighbors of a node. Clustering coefficient for a node $n$, having $k_n$ edges which connects it to $k_n$ neighbors is given below:
$$cc(n)=\frac{2*e_n}{k_n*(k_n-1)}$$

\textbf{Strength}~\cite{auber03} is another group level metric which extends the notion of calculating triads in a network. This metric quantifies the neighborhood's cohesion of a given edge and thus identifies if an edge is an intra-community or an inter-community edge. The \emph{strength} of an edge $e=(u,v)$, $s(e)$ is defined as follows:
 $$ str(e) = \frac{\gamma_{3,4}(e)}{\gamma_{max}(e)}$$
where $\gamma_{3,4}(e)$ is the number of cycles of size $3$ or $4$ the edge $e$ belongs to and $\gamma_{max}(e)$ is the maximum possible number of such cycles. Using this edge strength, one can define the \emph{strength} of a vertex as follows: 

$$str(v) = {\frac {\sum_{e \in adj(v)} str(e)}{deg(v)}}$$ 

where $adj(v)$ is the set of edges adjacent to $u$ and $deg(v)$ is the degree of $v$.
The idea is to quantify whether the neighbors of a node connect well to each other or are loosely connected to each other. The values range between [0,1] such that low values indicate poor connection whereas high values indicate strong connections among the neighbors of a node.

%---------------------------------- Burt's constraint -------------------------------- 
%\textbf{Burt's Constraint}~\cite{burt05} is a measure of a tie's strength and group cohesion. For a node v, it tells how much invested it is in people who are invested in other of v's neighbors. Mathematically, we can calculate it by using the following equation: 
%%$$Constr(v) =\sum_{j \in V_{i} \setminus\{i\}} (p_{ij} + \sum_{q \in V_{i} \setminus\{i,j\}}p_{iq}p_{qj} )^{2} $$
%The proportional tie strength is defined as 
%$$ p_{ij}=\frac{a_{ij}+a_{ji}}{**} $$
%One of the dominant methods is Ronald Burt's constraint measure, which taps into the role of tie strength and group cohesion. Another network based model is network transitivity.

\subsection{Network Level Centrality Metrics}
Network Level Metrics require the entire graph for calculation. Centrality in the context of network level metrics is a structure level metric which calculates how central a node is, in the entire network.

\textbf{Betweeness Centrality}~\cite{freeman77} calculates how often a node lies on the shortest path between any two pair of nodes in the network. Mathematically, the metric is defined as:
$$bc(v) = \sum_{u \neq v \neq w \in V} \frac{\mu_{uw}(v)}{\mu_{uw}}$$
where $\mu_{uw}(v)$ equals the number of shortest paths between two nodes $u$ and $w \in V$ going through the node $v$ and $\mu_{uw}$ equals the number of shortest paths between two nodes $u$ and $w \in V$.

\textbf{Eccentricity}~\cite{hage95} also tries to capture the notion of how central a node is in the network. The eccentricity ecc(v) of a node is the maximum distance between \textit{v} and any other node \textit{u} of G. Mathematically, eccentricity can be calculated by using the following equation:

$$ ecc(v)=\frac{1}{max\{d(v,u):u \in V \}}  $$

%Eccentricity represents the length of the chain of connections in a graph for a particular node. Its high value represents that the person has a long chain of relationship. He may not have many direct connections but he can reach to the person far from him in a graph through other connections. On the other side its low value (for e.g. ε(υ) = 1) shows that his friends have no more connections other than him.

\textbf{Closeness}~\cite{beauchamp65} is another network level metric which is the inverse sum of distances of a node to all other nodes given by the equation:
$$ clo(v)=\frac{1}{\sum \{d(v,u):u \in V \}}$$

% Social Interpretation: Its value represents the closeness of all the connections with the node. The smaller the value of closeness, the more closely are the connections with the node. Its max value can be 1 if it has single connection at a distance of 1. 

\begin{figure}[!t]
\centering
\includegraphics[width=0.45\textwidth]{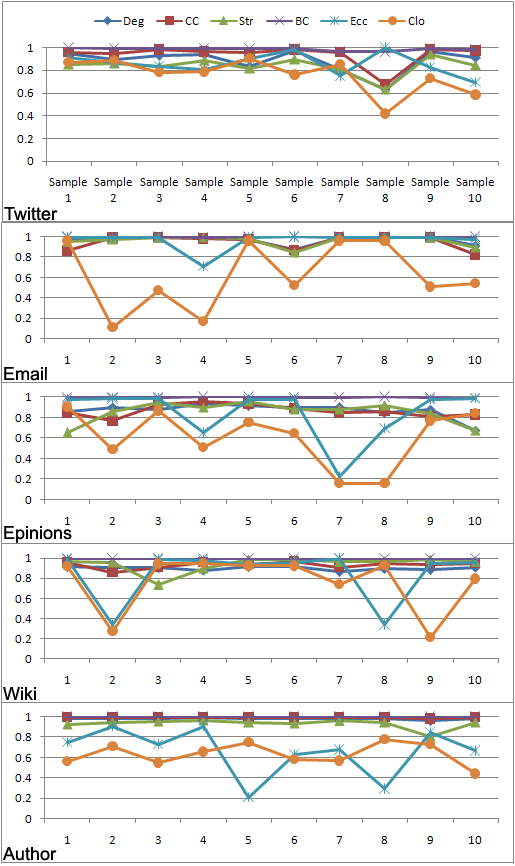}
\caption{Calculating different Network Metrics on each sample of the Five datasets.}
\label{fig::sampling}
\end{figure}

\section{Experimental Setup}\label{sec::experimental}

As the first step to perform a comparative analysis of various networks using different metrics, we perform sampling on all these data sets to obtain equal size networks in terms of number of nodes. In order to verify whether our samples truly reflect the original network we conducted a simple experiment which itself revealed interesting results about sampling methods. 

Three sampling methods exist in the literature for sampling graphs and networks. Node sampling, in which an induced sub graph of randomly drawn nodes is considered as a sample. Link sampling, in which randomly drawn edges are considered and their nodes are added to the network. And finally the most agreed upon sampling method for social networks~\cite{ahn07}, the snowball sampling which starting from a seed node, performs a breadth-first search collecting a subset of the entire graph \cite{rothenberg95}.
 
We used random repeated sampling collecting 10 samples of size 500 nodes from each data set giving us a total of 50 graphs. Next we calculated different network metrics on these samples. For each sample, we calculated the frequencies of the resulting values giving us a distribution of how these metric values occur in the network. For example, in case of the degree metric, we calculated the frequencies of the degree values obtained for the network. Next, for each data set we calculated the average of these frequencies. We then calculated the correlation coefficient of each sample from its average to give an idea of how much variation occurs in the sampled data.

Figure \ref{fig::sampling} clearly shows that closeness and eccentricity values for certain samples vary from the average values calculated for the respective data sets. For example closeness values for sample 2 and 4 for Email data, sample 2 and 9 for Epinions data, eccentricity values for sample 7 for Epinions, sample 8 for wiki and sample 5 and 8 for Author network have all very low values of correlation with the average values calculated for the respective samples. Both Eccentricity and Closeness are Network level metrics and represent how centric nodes are in a network. Eccentricity is the maximum distance a node can have from any other node, and Closeness is the average of the maximum distances from a node to all other nodes. Collecting a sample using snowball sampling is vulnerable with respect to both these metrics as the sampling method itself is based on generating paths from a seed node. 

\section{Inferences and Observations}\label{sec::inferences}

\begin{figure}[!t]
\centering
\includegraphics[width=0.5\textwidth]{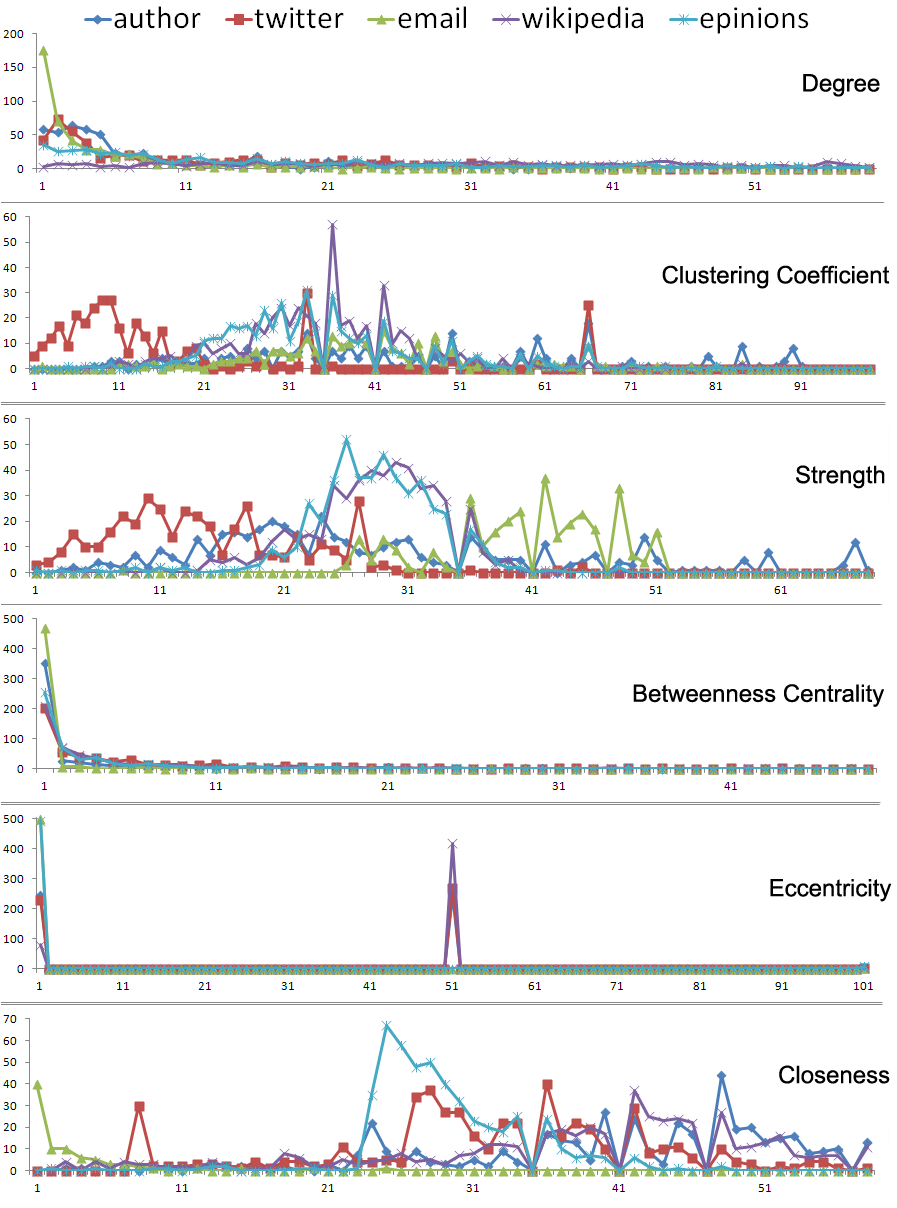}
\caption{Calculating different Network Metrics on the Five datasets. Horizontal axis represents bins and vertical axis represents the frequency with which nodes appear in that particular bin.}
\label{fig::results}
\end{figure}

Figure \ref{fig::results} shows the frequency distribution calculated for the above described metrics. These metrics either return values between 0 and 1, or have been normalized in this range to facilitate comparative study. Furthermore we have applied binning to calculate frequencies where the values have been rounded off to 2 decimal places giving us bins in the range $[0.00, 0.01, 0.02,\cdots,1.00]$. The values on the horizontal axis for the graphs in Figure \ref{fig::results} represent the bin number, i.e.\ bin 0 refers to the frequency of nodes for the value 0.00, bin 1 refers to the value 0.01 and so on. One final modification to these graphs is that we have cut the extreme bins for Degree distribution, Strength, Betweenness Centrality and Closeness as there was very less information available in these bins. 

From the degree distribution of the five networks in Figure~\ref{fig::results} the graphs for the author and the twitter network are quite similar. The most interesting observations are for the the wikipedia and the epinions network where we can see a linear decay in the degree distribution of the two networks which shows a non-scale free behavior of the two networks. The email network has a very high peak for very low values showing that most of the individuals in this network have used email very rarely for communication purposes.

The clustering coefficient frequencies have a similar behavior as all the networks have peaks in their frequency values. For example, the twitter network has a peak at bin 11 which refers to a value of $0.11$. This shows that around 30 nodes have a clustering coefficient of $0.11$. Other networks have a peak which starts from bin 21 to 51. The lowest peak is for the email network although the global clustering coefficient of this network is higher than other networks as shown in Table~\ref{tbl::stats}. 

A similar observation can be made about the frequencies for the strength metric as values gradually rise and fall off for every dataset. Wikipedia and epinions networks have frequencies quite close to each other, the email network has its peak shifted on the right and twitter's peak shifted on the left.

Betweenness centrality has the most perfect match for all these networks. This is due to a node with very high degree present in all networks, which in turn plays a central role in connecting short paths among pairs of nodes. These high degree values can be verified from Table\ref{tbl::stats}.

Eccentricity values of different networks follow each other very closely. This is again an implication of the presence a few very high degree nodes in the network as the maximum distance among any pair of nodes does not vary much, as all nodes use these high degree nodes which act as short cuts in these networks.

The most variation in the behavior of frequencies is for the closeness metric. The email network has initially high values as opposed to other networks but remains very low for other values. This is because of it has a node with exceptionally very high degree as it is connected to all other nodes. This reduces the average closeness of all pair of nodes. The twitter network has peaks around bin number 7, 27-28, 35 and 42 which is quite different from other networks. Wikipedia has also different peaks but they are shifted towards the right when compared to the twitter nework, which signifies higher frequencies for high closeness values. Epionions has a peak aournd bin 24 which gradually decreases and the author network has its peak value at around bin 46.

In general, the behavior of all these networks is similar when evaluated with the discussed metrics. Two findings can be quoted, one for the non-scale free behavior of two social networks, epinions and wikipedia. Second is the variations in frequencies for the closeness metric. 

\section{Conclusion}

In this paper, we have performed a comparative study to analyze contextually and semantically different social networks using different network statistics and metrics. Our results show that these network are structurally similar to each other in most of the cases.

We also demonstrated that snowball sampling method is vulnerable against two network level centrality metrics, eccentricity and closeness as repeated sampling on different data sets revealed inconsistent behavior of these networks.
As part of future work, we intend to incorporate more data sets and more network metrics to perform a comprehensive comparative analysis of different social networks. We also intend to explore the possibilities of proposing a new sampling method which is robust against different structural metrics.

\bibliographystyle{abbrv}
\bibliography{visu}

\end{document}